\documentclass[prl,showpacs,twocolumn]{revtex4}
\usepackage{graphicx,color}
\usepackage{amssymb}
\usepackage{rotating}
\usepackage{ dsfont }

\newcommand{\nn}{\nonumber}
\newcommand{\eqref}[1]{Eq.~(\ref{#1})}
\newcommand{\ket}[1]{\left|{#1}\right\rangle}
\newcommand{\bra}[1]{\left\langle{#1}\right|}
\newcommand{\scalar}[2]{\left\langle{#1}\middle|{#2}\right\rangle}

\def\be{\begin{equation}}
\def\ee{\end{equation}}
\def\bea{\begin{eqnarray}}
\def\eea{\end{eqnarray}}

\begin{document}

\title{Secular master equation for adiabatically time dependent systems}
\author{I. Kamleitner}
\affiliation{Institut f\"ur Theory der Kondensierten Materie, Karlsruher Institut f\"ur Technologie, 76128 Karlsruhe, Germany}
\begin{abstract}
  Relying only on first principles, we derive a master equation of Lindblad form generally applicable for adiabatically time dependent systems. Our analysis shows that the much debated secular approximation can be valid for slowly time dependent Hamiltonians when performed in an appropriate basis. We apply our approach to the well known Landau--Zener problem where we find that adiabaticity is improved by coupling to a low temperature environment.
\end{abstract}
\pacs{03.65.Yz, 03.67.-a, 75.50.Xx, 85.25.Cp} \maketitle


The adiabatic theorem~\cite{bor,mes} is an approximation derived in the very early days of quantum mechanics. Its application requires that the time dependence of an Hamiltonian $H(t)$ is slow compared to the internal time scales, as quantified by a small adiabatic parameter $\mathcal A$. The adiabatic theorem then states that the time dependence of the Hamiltonian only induces transitions between different instantaneous eigenstates of the order of $\mathcal A^2$. For an analytic Hamiltonian, this result can be improved~\cite{dav,ber2} to transitions of order of $e^{-1/\mathcal A}$.

The adiabatic theorem is of great significance in many traditional fields of quantum mechanics, like molecular physics, where it is a prerequisite for the use of the Born-Oppenheimer approximation~\cite{bor2}, and quantum optics where it is used, e.g.\ for stimulated Raman adiabatic passage~\cite{mil} and the related electromagnetically induced transparency~\cite{fle} to slow down light to near standstill. 

Renewed interest in the theorem came with Berry's discovery of geometric phases in cyclic adiabatic systems~\cite{ber, sim}. Berry's phase soon got generalized to non-commuting geometric rotations within a degenerate subspace, so called holonomic transformations~\cite{wil}. As these turned out to be robust to some errors in the control of experimental parameters, holonomies received a lot of attention in the context of quantum computing~\cite{zan,dua}. 

An entirely different approach to quantum computing, called adiabatic quantum computing~\cite{far, aha}, also heavily relies on the adiabatic theorem. The aim here is to find the ground state of a complicated Hamiltonian $H_{\rm final}$ of interacting qubits. This is achieved by first preparing the system in the ground state of an easy Hamiltonian $H_{\rm initial}$ (e.g.\ a non-interacting Hamiltonian). The Hamiltonian is then adiabatically changed to $H_{\rm final}$ and the adiabatic theorem guarantees that the final state is the desired ground state. This process can be used for quantum computing because it was shown that knowing the ground state of an appropriate Hamiltonian is equivalent to knowing the solution of the desired calculation.

The minimal time required for an adiabatic transformation is limited by $\mathcal A\ll 1$ and about an order of magnitude larger than typical non-adiabatic transformations. Even weak coupling to an environment might therefore significantly influence the dynamics of the system. In particular in respect to quantum computing, where high fidelity transformations are needed, decoherence is a major obstacle. It is of great importance to understand the interplay of decoherence and adiabaticity.

Adiabatic systems weakly coupled to an environment were already studied in the literature in connection to Berry's phase~\cite{eli,gam,sjo,kam,sar}, adiabatic quantum computing~\cite{chi,ami}, molecular magnets~\cite{rou}, biological systems~\cite{cai}, as well as Landau--Zener transitions~\cite{whi}. In some of these the authors postulated a Markovian master equation (ME) of Lindblad form similar to time independent systems. However, it is well known that for time dependent Hamiltonians the secular approximation can cause problems and hence a ME of Lindblad form has to be used with care. While~\cite{rou,cai} derived a Lindblad ME from first principles, they neglected terms of order $\mathcal A$ in the dissipative part of the ME. Although reasonable for the applications these authors had in mind, in the context of Cooper pair pumping it was shown that terms of order~$\mathcal A$ have to be included to find the correct transferred charge~\cite{sol,pek,kam2,sal}. In general, because the adiabatic theorem is correct up to $e^{-1/\mathcal A}$ it is desirable to derive a ME to the same order. This is done below.

The description of quantum systems by Lindblad MEs is standard for time independent systems~\cite{bre} and is desirable for two main reasons: First, quantum trajectory theory~\cite{bre} provides not only a neat physical intuition for dissipative processes, but it is also an efficient numerical method for solving Lindblad MEs by using a Monte Carlo algorithm, even for multi level systems. Second, it is the only ME to guarantee complete positivity~\cite{lin}. A crucial step to obtain a Lindblad ME is the secular approximation which neglects fast rotating terms. However, its validity for time dependent systems is not generally given and some work~\cite{whi,sol,pek,kam2,sal} uses non-secular MEs, which indeed result in negative probabilities for certain initial states. 

The aim of this work is to derive a completely positive ME for adiabatic, but otherwise very general systems (non--cyclic, multi--level), which is valid to exponential order $e^{-1/\mathcal A}$ in the adiabatic parameter. In particular, we rigorously justify the secular approximation leading to the desired Lindblad form. All other restrictions (weak coupling, short bath correlation time, separable initial state) are identical to the time independent case~\cite{bre}. Contrary to the time independent case, the Lindblad operators do not describe transitions between energy eigenstates, but transitions between super adiabatic states to be introduced below. The difference to the eigenstates is of order $\mathcal A$. While in some instances this modification results in rather small quantitative effects and might be neglected (as in~\cite{rou,cai,whi}), there are also situations (see our Landau-Zener example) where the super adiabatic states should be used, even for a correct qualitative understanding.

In adiabatic quantum computing, errors due to non-adiabaticities can be modelled by one or many successive Landau-Zener (LZ) transitions. Therefore, we apply our theory to the LZ problem where we find that pure dephasing does not alter the exponential decrease of the transition probability $P_{g\to e}=e^{-{\rm const}/v}$ with decreasing transition velocity~$v$, while the traditional Lindblad ME~\cite{rou,cai} wrongly predicts transitions of order $v^2$. Numerical solutions of our ME show that coupling to a low temperature environment even lowers the transition probability and stabilizes the adiabatic theorem.


We now derive the central result of this paper: A master equation of Lindblad form governing the evolution of the density operator of an adiabatically driven system weakly coupled to a Markovian bath. We take the Hamiltonian of system and bath of the form
\be
  H_{\rm{tot}} = H(t) + H_{\rm{b}} + A\otimes B_{\rm{b}},
\ee
where the subscript b refer to bath operators and no subscript refers to system operators. The generalization to a general interaction along the lines of~\cite{bre} is straight forward.  Assuming a factorizing initial density operator $\rho_{\rm{tot}}(0)=\rho(0)\otimes\rho_{\rm{b}}$ and a thermal state of the bath $\rho_{\rm{b}}\propto e^{-H_{\rm{b}}/k_BT}$, the systems density operator in the interaction picture satisfies the master equation~\cite{bre}
\bea
  \dot\rho(t) &=& \int_0^\infty d\tau\,\rm{Tr}[B(\tau)B\rho_{\rm b}]\nn\\
  &\times& [A(t-\tau) \rho(t) A(t)-A(t)A(t-\tau)\rho(t)] + \rm{h.c.}\qquad \label{eq2}
\eea
Here, the Born approximation $\rho_{\rm{tot}}(t)\approx\rho(t)\otimes\rho_{\rm{b}}$ and the Markov approximation $\rho(t-\tau)\approx\rho(t)$ have been employed which are valid for weak coupling to a bath with short memory time $\tau_b$; and h.c.\ denotes the hermitian conjugate. Operators with an argument are interaction picture operators, i.e.\ $A(t)=U^\dag(t,0)AU(t,0)$ where $U(t_2,t_1)=\mathcal T \exp[-i\int_{t_1}^{t_2}\! dt' H(t')]$ is the time evolution operator of the closed system and $\mathcal T$ denotes time ordering.

At each time we introduce a complete set of basis states $\ket{\phi_\alpha^t}$, which we will later choose such that decoherence acts in this basis, i.e.\ the environment causes the density operator to approach a diagonal state in this basis. Furthermore, we introduce
\bea
  \ket{\phi_\alpha^t(t-\tau)} &=& \exp\!\left[-i\!\int_{t-\tau}^t dt'\,E^t_\alpha(t')\right] U^\dag(t,t-\tau)\ket{\phi_\alpha^t} \quad \label{3}\\
  E_\alpha^t(t') &=& \bra{\phi^t_\alpha(t')}H(t')\ket{\phi^t_\alpha(t')},
\eea
which are, up to a phase factor, the states $\ket{\phi_\alpha^t}$ evolved backwards in time with $\ket{\phi_\alpha^t(t)}=\ket{\phi_\alpha^t}$. The phase factor in \eqref{3} is chosen such that the states are \emph{parallel transported} $\bra{\phi_\alpha^t(t-\tau)}\frac{d}{d\tau}\ket{\phi_\alpha^t(t-\tau)}=0$ to minimize their rate of change with $\tau$. We neglect the change of the energies within $\tau_b$ and use $\exp\!\left[i\int_{t-\tau}^t dt'\,E^t_\alpha(t')\right]\approx e^{iE_\alpha^t\tau}$. Although our main results are valid without this approximation, it is not a strong restriction for adiabatic systems and significantly simplifies the notations.

Inserting twice the identity operator $\sum_\alpha \ket{\phi_\alpha^t}\!\bra{\phi_\alpha^t}$ for $A(t)$ as well as $\sum_\alpha \ket{\phi_\alpha^t(t-\tau)}\!\bra{\phi_\alpha^t(t-\tau)}$ for $A(t-\tau)$ we get
\bea
  \dot\rho(t) &=& \sum_{\alpha\alpha'\beta\beta'} \int_0^\infty d\tau\,\rm{Tr}[B(\tau)B\rho_{\rm b}]\, e^{i\omega^t_{\beta\beta'}\tau} \nn\\
  &\times& \bra{\phi_\beta^t(t-\tau)}A\ket{\phi_{\beta'}^t(t-\tau)} \bra{\phi_{\alpha'}^t}A\ket{\phi_\alpha^t} \nn\\
  &\times& \!\left[ U^\dag(t,0)\!\ket{\phi_\beta^t}\!\bra{\phi_{\beta'}^t}\!U(t,0) \rho(t) U^\dag(t,0)\!\ket{\phi_{\alpha'}^t}\!\bra{\phi_\alpha^t}\!U(t,0) \right.\! \nn\\
  && - \left. \delta_{\alpha\beta}U^\dag(t,0)\ket{\phi_{\alpha'}^t}\!\bra{\phi_{\beta'}^t}U(t,0) \rho(t) \right]+ \rm{h.c.}, \label{5}
\eea
where we introduced the transition frequencies $\omega^t_{\beta\beta'}=(E_{\beta'}^t-E_\beta^t)$.

At this stage we ask the question: Into which state does the system relax if the bath has zero temperature, i.e.\ if Tr$[B(\tau)B\rho_B]$ has only positive frequencies? We can find a definite answer if $\big\langle\phi_\beta^t(t-\tau)\big|A\big|\phi_{\beta'}^t(t-\tau)\big\rangle$ has only frequencies in $\tau$ which are small compared to $|\omega_{\beta\beta'}^t|$. Then, the $\tau$-integral of \eqref{5} vanishes for $E_\beta^t>E_{\beta'}^t$ and transitions $\beta \leftrightarrow \beta'$ only go towards lower energy. Therefore, the system evolves towards the state $U^\dag(t,0)\big|\phi_\beta^t\big\rangle$ (or $\big|\phi_\beta^t\big\rangle$ in the Schr\"odinger picture) with lowest energy, similar to an open quantum system with a time independent Hamiltonian. It is important to realize that this conclusion holds only if we can find a basis $\ket{\phi_\alpha^t}$ such that $\big\langle\phi_\beta^t(t-\tau)\big|A\big|\phi_{\beta'}^t(t-\tau)\big\rangle$ with $\ket{\phi_\alpha^t(t-\tau)}$ as defined in \eqref{3} does not oscillate with fast frequencies. 

For a time independent Hamiltonian one can trivially find a basis to satisfy the above statement, namely the eigenstates of the Hamiltonian. For adiabatic systems we can also find a basis which upon application of the exact time evolution operator changes only with frequencies much smaller than the transition frequencies, the so--called super adiabatic basis described below.

\begin{figure}[t]
	\includegraphics[width=0.95\linewidth,height=1.2\linewidth]{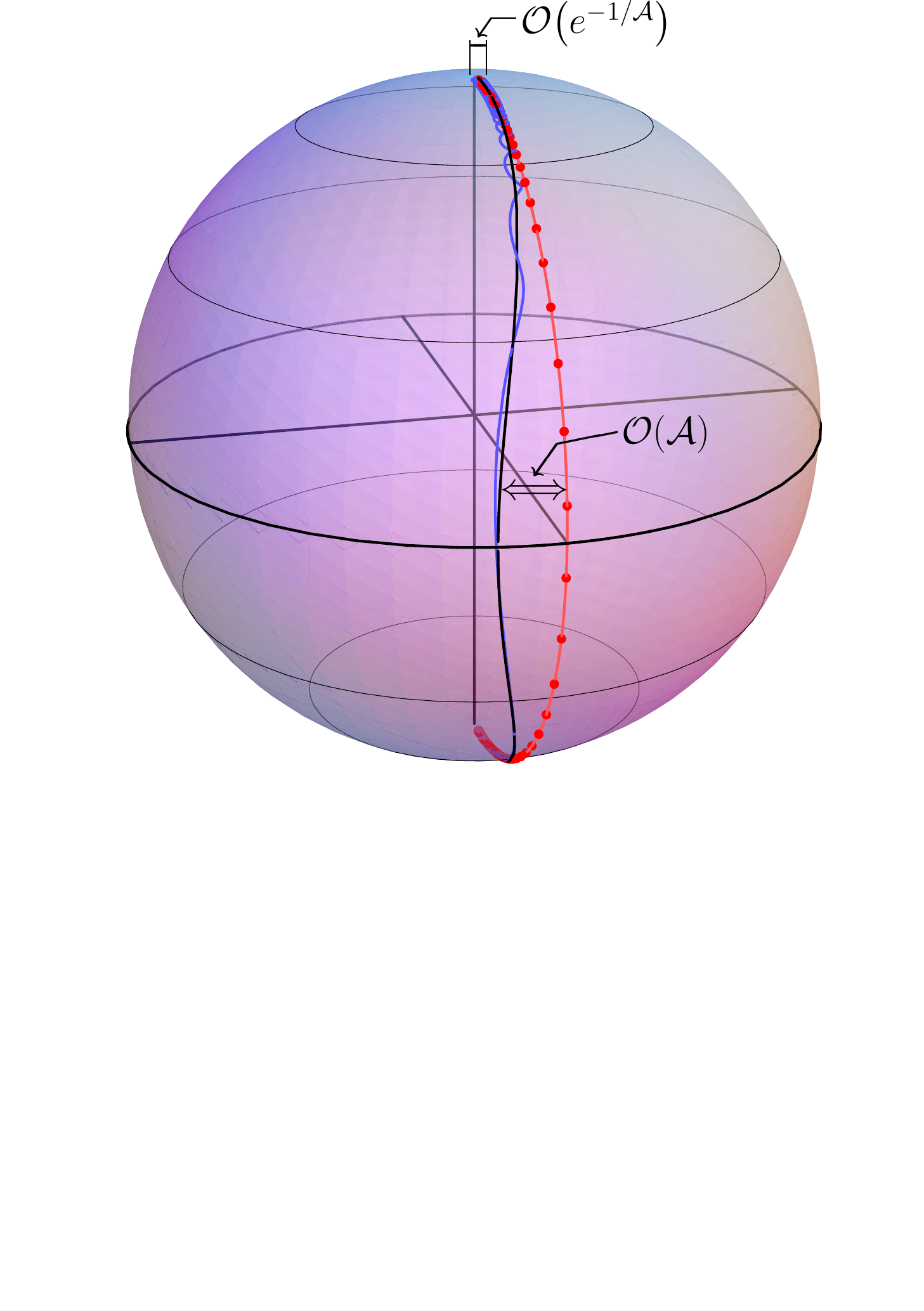}\vspace{-41mm}
	\caption{(Color online) The red line is the Bloch vector of the instantaneous ground state of the LZ problem and the dots are equidistant time steps. The Black curve is the third order super adiabatic ground state and the blue curve is the actual time evolution of an initial ground state. The time evolution operator causes oscillations around the super adiabatic ground state. One result of this paper is that a low temperature environment causes the system to relax onto the black curve, rather than the red one, therefore stabilizing the adiabatic theorem. \label{fig1}}
\end{figure}

If the system at some time $t$ is in the instantaneous eigenstate $\ket{n_\alpha(t)}$, then the time evolution operator $U^\dag(t,t-\tau)$ will cause the state to oscillate with the transition frequencies and with an amplitude of order of the adiabatic parameter $\mathcal A(t)$ defined by
\be
  \mathcal A = \max_{\alpha\beta}\frac{|\scalar{n_\alpha(t)}{\dot n_\beta(t)}|}{|\varepsilon_\alpha(t)-\varepsilon_\beta(t)|},
\ee
where $\varepsilon_\alpha(t)$ are the instantaneous eigenenergies of $H(t)$ and the dot denotes the time derivative. If the system is in an eigenstate $\big|n^{1}_\alpha(t)\big\rangle$ of the first order super adiabatic Hamiltonian $H_1=H+i\dot R_1R_1^\dag$, then the oscillation amplitude is only of order $\mathcal A^{2}$. The operator $R_1=\sum_\alpha\ket{n_\alpha(0)}\!\bra{n_\alpha(t)}$ transforms the system into the instantaneous basis. According to~\cite{ber2}, one can use the $j$-th order super adiabatic basis $\big|n^{j}_\alpha(t)\big\rangle$ to further lower the oscillation amplitude to order $\mathcal A^{j+1}$. However, once $j\approx1/\mathcal A$ the oscillation amplitude starts to diverge as $j!$; with the minimal amplitude of order $e^{-1/\mathcal A}$ if the Hamiltonian is analytic in time. An example is shown in Fig.~\ref{fig1}.

Therefore, in the following we make the explicit choice $\ket{\phi_\alpha^t}=\big|n^{j}_\alpha(t)\big\rangle$ where $j$ is the natural number nearest to $1/\mathcal A$ and for adiabatic systems we can safely neglect the exponentially small oscillations caused by the evolution operator~\cite{footnote1,footnote0}. The system will relax not into the instantaneous ground state, but into the super adiabatic ground state! We will see later that for LZ transitions this result is of importance for certain regimes.

It is well known that the master equations~(\ref{eq2}) and~(\ref{5}) do not preserve positivity~\cite{bre}. The most common way to fix this is to perform a secular approximation which drops the terms in the interaction picture ME which rotate fast in $t$, as they average to zero. (Another method~\cite{whi2} uses a $\tau$-integral only from zero to $t$). The secular approximation might give problems for time--dependent Hamiltonians as many more frequencies become involved, some of them not large enough to justify the secular approximation. However, according to~\cite{dav,ber2}  $U^\dag(t,0)\ket{\phi_\alpha^t}=\ket{\phi_\alpha^0} \exp\!\left[i\int_0^t dt'\,E_\alpha^{t'}\right]+\mathcal O\left(e^{-1/\mathcal A}\right)$ and upon neglecting the exponentially small non--adiabaticities, we are only left with the frequencies $E_\alpha^t=\bra{\phi_\alpha^t}H(t)\ket{\phi_\alpha^t}$, much like in the time independent case. We conclude that the secular approximation is justified for adiabatic systems if performed in the super adiabatic basis. Our final Schr\"odinger picture master equation of Lindblad form reads~\cite{footnote2}
\begin{eqnarray}
	\dot\rho &\!=\!& -i[H+H_{\rm{LS}},\rho] + L_{0}\rho L^\dag_{0}-\frac12 \big\{L^\dag_{0}L_{0},\rho \big\}  \nn\\
	&&+ \sum_{\alpha\neq\beta}  L_{\alpha\beta}\rho L^\dag_{\alpha\beta}-\frac12 \big\{ L^\dag_{\alpha\beta}L_{\alpha\beta},\rho \big\}  \label{Lindblad}
\end{eqnarray}
with time dependent Lindblad operators
\begin{eqnarray}
	L_{0}(t) &=& \sqrt{\gamma(0)} \sum_\alpha \bra{\phi_\alpha^t}A\ket{\phi_\alpha^t} \ket{\phi_\alpha^t}\!\bra{\phi_\alpha^t},  \qquad \\
	L_{\alpha\beta}(t) &=& \sqrt{\gamma(\omega_{\alpha\beta}^t)} \bra{\phi_\alpha^t}A\ket{\phi_\beta^t}  \ket{\phi_\alpha^t}\!\bra{\phi_\beta^t}. \label{L}
\end{eqnarray}
and the Lamb shift Hamiltonian $H_{\rm{LS}}(t)=\sum_{\alpha\beta}S(\omega_{\alpha\beta}^t) |\!\bra{\phi_\alpha^t}A|\phi_\beta^t\rangle|^2 \big|\phi_\beta^t\big\rangle\!\big\langle\phi_\beta^t\big|$. Here, $\gamma(\omega)$ and $S(\omega)$ are the real and imaginary parts of the Laplace transform of the bath correlation function Tr$[B(\tau)B\rho_b]$, respectively. We note that neither Lamb shift $H_{\rm{LS}}$, nor dephasing $L_0$ affects the system if it is in a super adiabatic state, and that relaxations and excitations $L_{\alpha\beta}$ change the populations of the super adiabatic states.

A note on the time scales: Like all secular weak coupling master equations, \eqref{Lindblad}~-~(\ref{L}) are valid provided the coupling is weak compared to the inverse bath correlation time~$\tau_b^{-1}$ and that relaxation and dephasing times are long compared to the energy differences $\varepsilon_\alpha(t)-\varepsilon_\beta(t)$. Adiabaticity requires $|\!\scalar{n_\alpha(t)}{\dot n_\beta(t)}\!| \ll |\varepsilon_\alpha(t)-\varepsilon_\beta(t)|$ and we also used $|\!\scalar{n_\alpha(t)}{\dot n_\beta(t)}\!|\ll\tau_b^{-1} $. We emphasize that we do not assume $\tau_b^{-1}\gg|\varepsilon_\alpha(t)-\varepsilon_\beta(t)|$, which would be valid only for extremely fast vanishing bath correlations.


We now go on to apply our master equation \eqref{Lindblad} to the Landau-Zener problem with
\bea
  H(t) &=& \frac12 \left( \begin{array}{cc} -vt&\Delta \\ \Delta&vt \end{array} \right). \label{eq10}
\eea
Without coupling to a bath, the transition probability from the ground state at $t=-\infty$ to the excited state at $t=\infty$ was found~\cite{lan,zen,stu,maj} to be $P_{g\to e}=e^{-\pi\Delta^2/(2v)}$. Although the sweep from $-\infty$ to $\infty$ is not realistic, finite sweeps through an avoided level crossing which are performed in many experiments are very well approximated by this transition formula.

The coupling to the environment is via $A=\sigma_z$, such that relaxation processes vanish in the limit $t\to\pm\infty$. In the case of a bosonic bath, there are very nice exact~\cite{ao,wub} and numerically exact~\cite{ort} results, which show that a $\sigma_z$ coupling to a zero temperature bath does not modify $P_{g\to e}$. The reason is an astonishing cancellation of amplitudes which move into excited bath states at different times. This strictly non-Markovian effect can not be captured by a master equation technique applied here or in~\cite{whi}. However, this cancellation of amplitudes requires that the temperature of the bath is exactly zero in a sense that all bath modes are in the ground state. While it is realistic in many experimental setups to achieve temperatures small compared to the level splitting of the system, a temperature below the slowest bath modes is much more challenging~\cite{footnote}. Therefore, a Markovian weak coupling calculation might reveal interesting physics not captured by exact calculations, as long as the bath correlation time is short (which is not the case for exactly zero temperature).

In many solid state experiments the Hamiltonian can only be obtained qualitatively from first principles, while the precise level splittings are obtained experimentally and already include the Lamb--shift. For this reason we will discard $H_{\rm{LS}}$, although for some applications involving the use of a LZ transition to characterize the environment~\cite{whi}, the Lamb--shift is of great importance.

\begin{figure}[t]
	\includegraphics[width=1\linewidth]{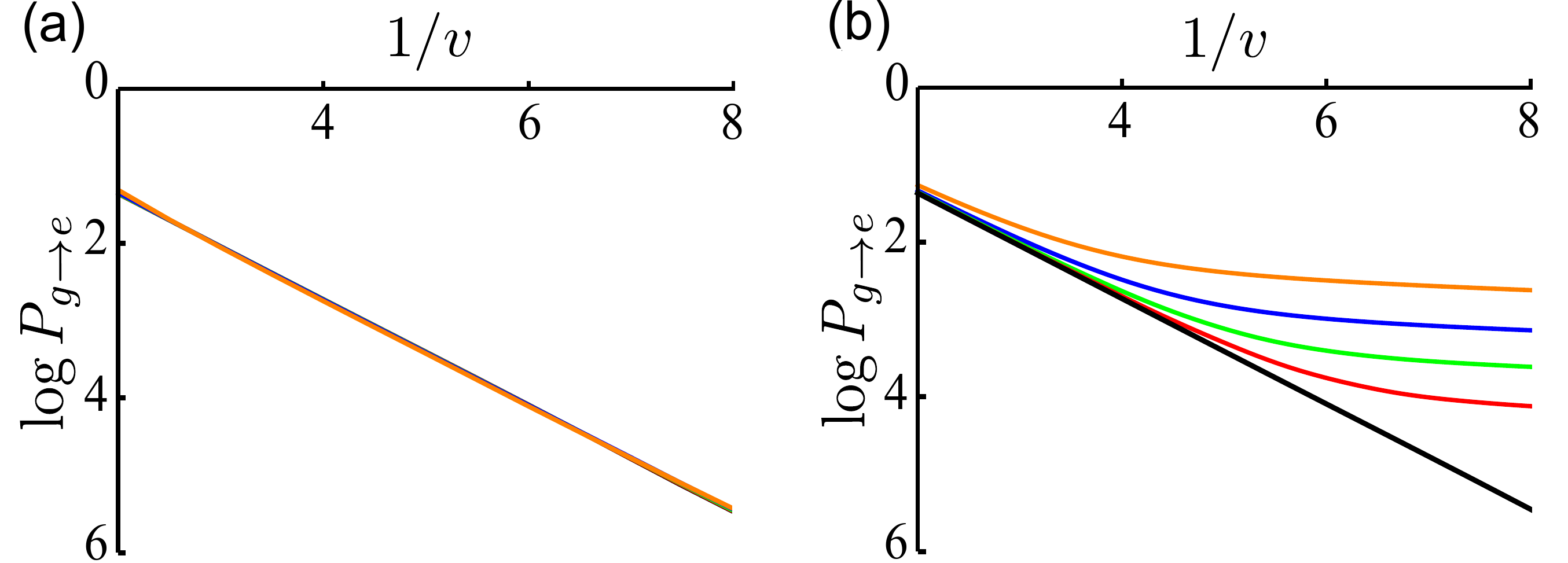}
	\caption{(Color online) The transition probability $P_{g\to e}$ for $\Delta=1$ as a function of $1/v$ without coupling (black), and with dephasive coupling $\gamma(0)=0.003$~(red), 0.01~(green), 0.03~(blue), and 0.1~(orange). In panel~(a) we used \eqref{Lindblad}~-~(\ref{L}) and the forth order super adiabatic states for $\ket{\phi_\alpha^t}$, while in panel~(b) we used the instantaneous eigenstates of $H(t)$.  \label{fig2}}
\end{figure}

We obtained the transition probability $P_{g\to e}$ by numerically solving \eqref{Lindblad}, and plots versus the inverse transition velocity $v^{-1}$ are shown in Fig.~\ref{fig2}~(a) for different coupling strength to a slow environment, i.e.\ for pure dephasive coupling. The fact that all lines fall on top of each other show that dephasing has no effect on LZ transitions. That is to be expected as a constant (or slow) environment coupled via $\sigma_z$ merely shifts the time of the LZ transition. If we used the traditional ME~\cite{rou,cai} with Lindblad operators given in terms of instantaneous eigenstates, then we would not find an exponentially small $P_{g\to e}$, but only a quadratically small one, as shown in Fig.~\ref{fig2}~(b). This can be understood by using quantum trajectory theory~\cite{bre}. A single trajectory continuously evolves along the blue line of Fig.~\ref{fig1} until a quantum jump rotates the state around the instantaneous ground state (red line). Subsequent continuous Hamiltonian evolution leads to oscillations around the super adiabatic state (black curve). Averaging over different jump times lead to a population of the excited state of order of $\mathcal A^{-2}$. The same argument also explains that pure dephasing does not alter the exponentially small transition amplitudes if the correct Lindblad operators are used.

\begin{figure}[t]
	\includegraphics[width=1\linewidth]{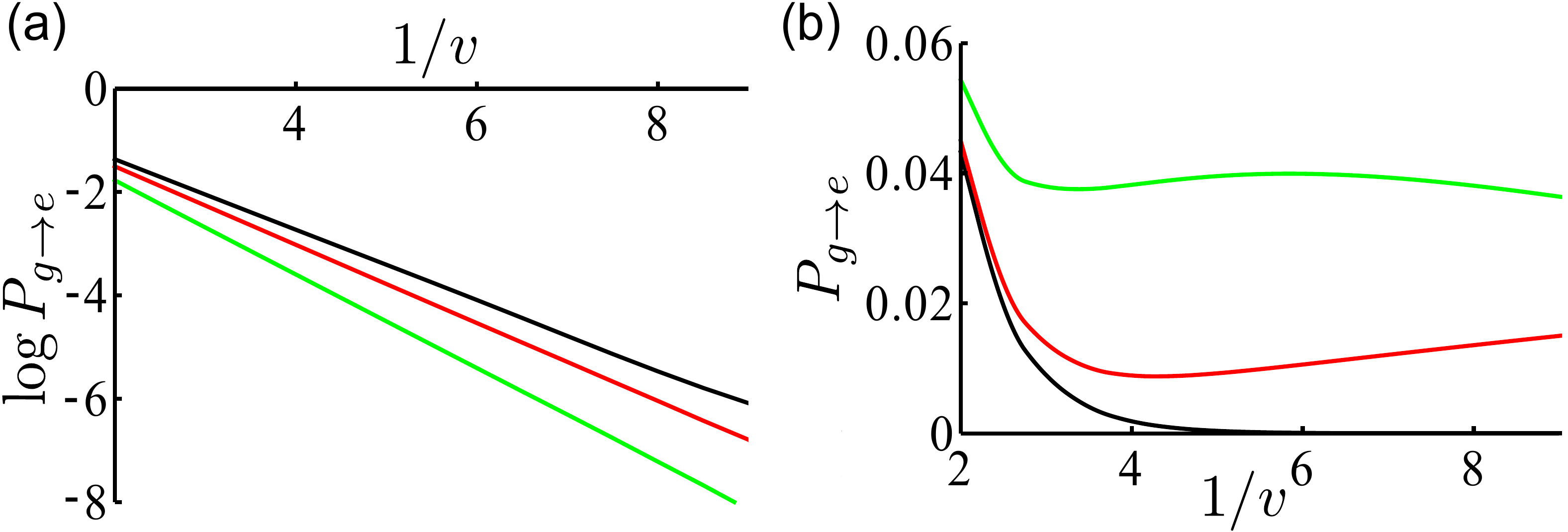}
	\caption{(Color online) The transition probability $P_{g\to e}$ for $\Delta=1$ as a function of $1/v$ for an Ohmic environment. Panel~(a) is for a low temperature environment $T\ll\Delta$ with $\gamma_0=0$~(black), 0.01~(red), and 0.03~(green). For panel~(b) we used $T=0.5$ and $\gamma_0=0$~(black), 0.01~(red), and 0.1~(green). \label{fig3}}
\end{figure}

To discuss relaxation and excitation processes, we choose an ohmic bath correlation function
\be
  \gamma(\omega) = \frac{\gamma_0 \omega e^{-\omega/\omega_c}}{1-e^{-\omega/k_BT}}
\ee
with a cut--off frequency $\omega_c=5\Delta$. Fig.~\ref{fig3}~(a) shows the transition probability for a low temperature bath for various coupling strengths $\gamma_0$. It can be seen that increasing the coupling is equivalent to decreasing the velocity $v$. The result that a low temperature bath can stabilize adiabatic ground state evolution is already known from Cooper pair pumping~\cite{pek}, but contradicts the exact spin--boson calculations~\cite{ao,wub}. This is a manifestation of the reasoning described three paragraphs earlier.

The finite temperature transition probability is plotted in Fig.~\ref{fig3}~(b). First, it decreases with $1/v$ due to improving adiabaticity, and then increases because the environment has more time to excite the system. For even slower velocities, in particular for stronger coupling, $P_{g\to e}$ decreases again, because the system has more time to relax at times when $vt\gg T$.


To summarize, we used the same premises which are generally used to derive a master equation of Lindblad form for time independent Hamiltonians, and extended the analyses to adiabatically time dependent Hamiltonians. Our treatment shows that the validity of the secular approximation can be extended to adiabatic systems. We further found that the system decoheres into the super adiabatic states in the sense that off--diagonals of the density matrix diminish in this basis due to dephasing, and that the system tends to relax to the super adiabatic ground state if the environment has low temperature. We applied our ME to the Landau-Zener problem where we found intuitive results while we also showed that a more naive ME with decoherence in the instantaneous eigenstates of the Hamiltonian would lead to wrong transition probabilities. A positive result is that coupling to a low temperature bath can stabilize adiabaticity.

\emph{Note:} During the writing of this letter, a derivation of a quite general adiabatic ME with secular approximation appeared~\cite{alb}, which describes approximations similar to here in much detail and is therefore of great value. The main difference to our work is that they assume $U^\dag(t,t-\tau)\approx e^{i\tau H(t)}$, similar to~\cite{whi}, and therefore do not use super adiabatic states in the Lindblad operators. As is clear from their and our work, this neglects contributions of order $\gamma_0^2 \mathcal A$ (unless $\tau_b^{-1}\gg \omega_{\alpha\beta}$).

\acknowledgements{The author is very thankful for stimulating discussions with P. Orth, R. Whitney, and A. Shnirman. This work was funded by the EU FP7 GEOMDISS project.}


\begin{thebibliography}{99}
\bibitem{footnote0} High order superadiabatic states often have rather large analytical expressions. Depending on the required precision one might be satisfied with some lower order super adiabatic state. E.g., first order states are sufficient for a qualitative understanding of Cooper pair pumps.
\bibitem{footnote1} The Floquet states of Ref.~\cite{kam2}, which for cyclic adiabatic systems are very similar to the super adiabatic states, also oscillate with an amplitude of order $e^{-1/\mathcal A}$. The advantages of the $n$-th order superadiabatic states are that they apply also for non-cyclic Hamiltonians, and that at least for two-level systems, they can be obtained analytically.
\bibitem{footnote2} We approximate $\bra{\phi_\beta^t(t-\tau)}A\ket{\phi_{\beta'}^t(t-\tau)}\approx\bra{\phi_\beta^t(t)}A\ket{\phi_{\beta'}^t(t)}$. This is allowed because the term does not introduce new significant frequencies, and $\ket{\phi_{\beta'}^t(t-\tau)}$ changes little for $\tau\lesssim\tau_b$. This approximation, which is valid if $\mathcal A \omega_{\alpha\beta}\tau_b\ll1$, can in principle be avoided by absorbing the $\tau$--dependence into an effective bath correlation function.
\bibitem{footnote} Furthermore, realistic bosonic bath modes will decohere themselves. Then it is not clear if amplitudes acquired by the bath at different times still cancel, or if they add up incoherently.
\end{thebibliography}
\end{document}